# Polarization Weight Family Methods for Polar Code Construction

Yue Zhou, Rong Li, Huazi Zhang, Hejia Luo, and Jun Wang
Huawei Technologies, Co. Ltd., Hangzhou, China
Email:{zhouyue5, lirongone.li, zhanghuazi, luohejia, justin.wangjun}@huawei.com

*Abstract*—**Polar codes are the first proven capacity-achieving codes. Recently, they are adopted as the channel coding scheme for 5G due to their superior performance. A polar code for encoding length-$K$ information bits in length-$N$ codeword could be specified by the polar code construction method. Most construction methods define a polar code related to channel parameter set, e.g. designed signal-to-noise ratio. Polarization weight (PW) is a channel-independent approximation method, which estimates the subchannel reliability as a function of its index. In this paper, we generalize the PW method by including higher-order bases or extended bases. The proposed methods have robust performance while preserving the computational and mathematical simplicity as PW.**

## I. Introduction

Recently, polar codes have been identified as one of the channel coding schemes in the 5G enhanced Mobile Broad Band (eMBB) due to their excellent performance [1].

Arikan firstly proved that polar codes could achieve the capacity of any symmetric binary input symmetric discrete memoryless channels (B-DMCs) under a successive cancellation (SC) decoder as the code length goes to infinity [2]. From then on, Polar codes have drawn increasing research attention.

For an $(N, K)$ Polar code, the selection of the information set $\mathcal{I}$ according to synthesized channel reliability is referred to as Polar construction [2]. To select $K$ good channels out of the $N$ sub-channels, an ordering of sub-channel reliabilities should be estimated. For binary erasure channels (BEC), Bhattacharyya parameter was used as the reliability metric [2]. For other channels, Mori and Tanaka [3] use density evolution (DE) for a more accurate reliability evaluation, but suffers from excessive complexity. Tal and Vardy proposed to reduce complexity with a close-to-optimal quantizer [4]–[7]. Afterwards, Gaussian approximation (GA) was proposed to further reduce the computational complexity of DE [8]. GA uses a two-segment approximation function to cut-off the computational complexity when applied to binary input additive white Gaussian noise (AWGN) channels, but yielding almost the same precision.

In coding theory, most of the codes are constructed independent of the channel signal-to-noise ratio (SNR), however Polar codes typically defines the set $\mathcal{I}$ of polar codes with a given channel which minimizes the block error rate (BLER) under SC decoder. Recently, a Polar code construction method called polarization weight (PW) was proposed, which gives the reliability ordering as a function of their indices [9]. The PW method shows stable performance with a low computational cost [10].

In this paper, we propose to generalize the PW method in order to obtain even better performance. Two methods, named higher-order PW (HPW) and extended PW (EPW), are elaborated. The numerical simulation results show that HPW and EPW could approach the optimal performance under a SC decoder, achieving robust performance under list decoders. The rest of this paper is organized as follows. The preliminaries of polar encoding and decoding are firstly described in Section II. Then, the HPW and EPW methods are proposed in Section III. A comprehensive performance comparison of HPW, EPW and GA is investigated in Section IV. Finally, Section V. summarizes the paper.

## II. Preliminaries

### A. Notations

The calligraphic characters, such as $\mathcal{X}$, denote sets and $|\mathcal{X}|$ denotes cardinality of $\mathcal{X}$. Lowercase letters (e.g., $x$) denote scalars. $\mathbf{v}_i^N$ denotes a vector $(v_i, v_2, \ldots, v_N)$, and $\mathbf{v}_i^j$ denotes a subvector $(v_i, v_{i+1}, \ldots, v_j)$. The sets of binary and integer field are denoted by $\mathbb{B}$ and $\mathbb{Z}$, respectively. Only square matrices are considered with polar coding, which are denoted by bold letters. The subscript of a matrix indicates its size, e.g. $\mathbf{F}_N$ represents an $N \times N$ matrix $\mathbf{F}$. The submatrix formed by the rows with indices in $\mathcal{I}$ are denoted as $\mathbf{F}(\mathcal{I})$. The modulo-2 operation between two matrices $\mathbf{F}$ and $\mathbf{G}$ is expressed as $\mathbf{F} \oplus \mathbf{G}$, and the $n$-folded Kronecker power of $\mathbf{F}$ is denoted by $\mathbf{F}^{\otimes n}$.

### B. Polar Codes Encoding and Decoding

Polar codes are rooted in the channel polarization phenomenon. At first, the same independent channels are transformed into two kinds of synthesized sub-channels: more reliable channels and less reliable channels. By recursively applying such polarization transformation, when the code length is sufficient, the synthesized sub-channels converge to two extreme groups: the noisy sub-channels and almost noise-free sub-channels. Since the noiseless channels have higher capacities/reliabilities than the noisy channels, polar codes transmit information bits over the noiseless sub-channels while assigning frozen bits (fixed value of zeroes or ones, and assumed known at both the encoder and decoder) to the noisy ones.

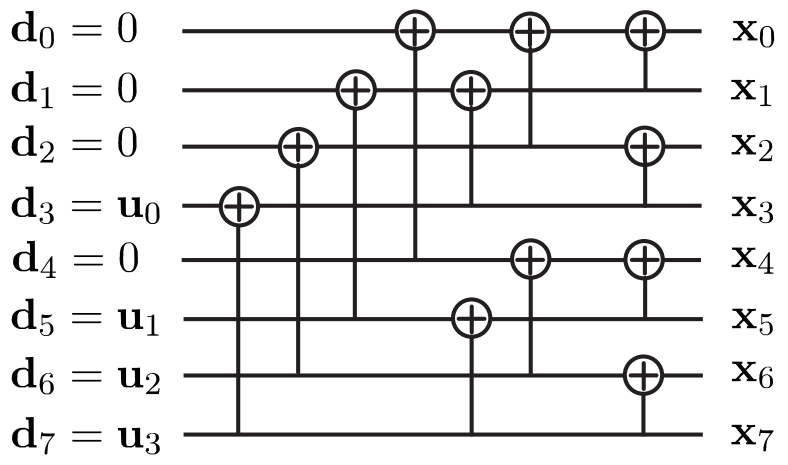

Fig. 1. Encoding implementation with $(N, K, \mathcal{I}) = (8, 4, \{3, 5, 6, 7\})$

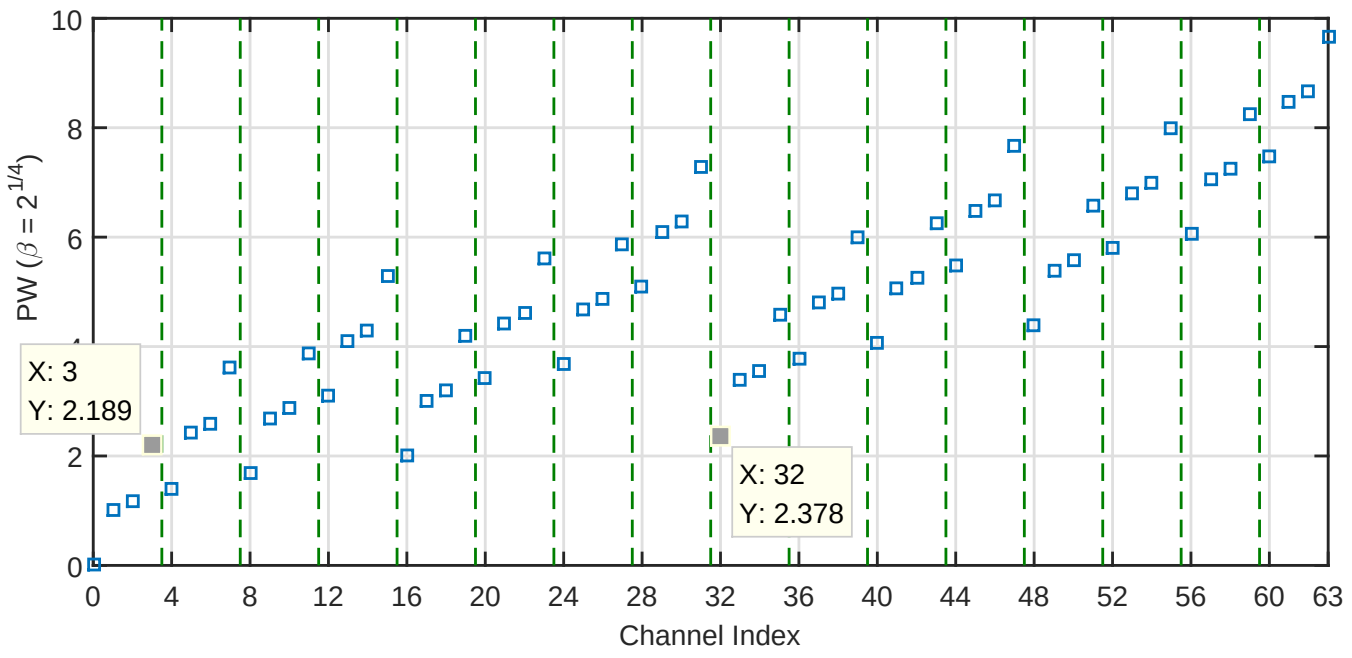

Fig. 2. The PW value of each sub-channels for a codeword length $N = 64$ polar code

However, for finite code length $N$, the polarization of the sub-channels is incomplete. Sub-channels with different reliabilities are in between the noiseless (high reliability) sub-channels and noisy (low reliability) sub-channels. To choose a subset $\mathcal{I}$ of $K$ sub-channels from $\{0, 1, \dots, N-1\}$ to encode $K$ information bits becomes the polar code construction problem. Here $N$ is restricted to powers of two ($N = 2^n$, $n \geq 0$), and the rest of $\mathcal{I}$ in $\{0, 1, \dots, N-1\}$ is called frozen sub-channels which denoted as $\mathcal{F}$. Thus, a polar code can be specified completely by $(N, K, \mathcal{I})$ and the corresponding encoding procedure can be described as

$$\mathbf{x}_0^{N-1} = \mathbf{d}_0^{N-1}\mathbf{G}_N, \tag{1}$$

where $\mathbf{d} \in \mathbb{B}^N$ and $\mathbf{G}_N$ is the generator matrix of order $N$, defined as

$$\mathbf{G}_N = \mathbf{F}^{\otimes n} \tag{2}$$

with the Arikan's standard polarizing kernel $\mathbf{F} \triangleq \left[\begin{smallmatrix} 1 & 1 \\ 0 & 1 \end{smallmatrix}\right]$.

Considering the information and frozen set, we may write (1) as

$$\mathbf{x}_0^{N-1} = \mathbf{d}_\mathcal{I}\mathbf{G}_N(\mathcal{I}) \oplus \mathbf{d}_\mathcal{F}\mathbf{G}_N(\mathcal{F}). \tag{3}$$

We consider the frozen bits as zeroes, $\mathbf{d}_\mathcal{F} = \mathbf{0}$, and the information bits as the information to be encoded, $\mathbf{d}_\mathcal{I} = \mathbf{u}$. Fig. 1 illustrates an encoding example with polar code $(8, 4, \{3, 5, 6, 7\})$.

The SC decoding algorithm can be regarded as a greedy search algorithm over the compact-stage code tree. Between the two branches associated with an information bit at a certain level, only the one with the larger probability is selected for further processing. Whenever a bit is wrongly determined, correcting it in future decoding procedures becomes impossible.

As an enhanced version of SC, the SC List (SCL) decoder [11] searches the code tree level by level, in much the same manner as SC. However, SCL allows a maximum of $L$ candidate paths to be further explored, which preserves the further error correction ability. Cyclic redundancy check (CRC)-aided SCL decoding scheme is a kind of SCL decoder, which outputs the SCL candidate paths into a CRC detector, and the check results are utilized to detect the correct codeword [11], [12].

Most construction methods are designed to choose the information set $\mathcal{I}$ which optimizes the BLER performance of a SC decoder. For practical implementation, list decoders with varied list sizes are widely used. The optimal construction methods for a SC decoder may not work well under SCL/CA-SCL decoders. Thus, a construction method suitable for various list size decoder is required.

## III. Polarization Weight Family Method

PW was first proposed in 3GPP RAN1 #86 as a method to generate an ordered sequence of sub-channels by reliability [9]. The SNR-independent sub-channel reliability order is estimated by computing the PW of each sub-channel and storing the ordered index sequence $\mathbf{q}_0^{N_{\max}-1}$ for the polar code of maximum code length $N_{\max}$. The PW of each sub-channels is defined as a weighted summation in the binary domain of the corresponding subchannel index. Assuming function $T(i) \triangleq B_{n-1}B_{n-2}...B_0$ transforms decimal channel index $i$ into $n$ bit binaries with the most significant bit on the left and $T(i,j) \triangleq B_j$, where $i \in \mathbb{Z}$, $B_j \in \mathbb{B}$, $j = [0, 1, \dots, n-1]$ and $n = \log_2(N_{\max})$. Then, the PW of sub-channel $i$ is

$$W_i = \sum_{j=0}^{n-1} T(i,j) * \beta^j = \sum_{j=0}^{n-1} B_j * \beta^j, \tag{4}$$

where the $\beta$ is a constant weight base in the summation. This mathematical formula was regarded as $\beta$-expansion [10], which represents the polarization weight of a sub-channel $W_i$ with the base $\beta$ through the summation of (4).

From (4), PW provides a channel-parameter-free polar code construction method in a neat mathematical equation which only involves a single input variable — channel index. However, the chosen of the base $\beta$ will affect the PW of each sub-channel and then the ordered sequence. Thus, the choice of the representing base $\beta$ defines the performance of PW methods. At first, the base $\beta$ was recommend as $\beta = 2^{1/4}$ in [9]. Then, the interval for base $\beta$ was proven to converge to a constant close to $1.1892 \approx 2^{1/4}$, through the universal partial order theory and the comparison with the DE/GA generated sequence order for AWGN channels [10].

Choosing the base $\beta = 2^{1/4}$, the PW method has the same performance compared with GA under SCL decoder with list size 8 [10]. However, for varied list size SCL decoders, PW may not provide the best performance. To keep the simplicity of the mathematical representation, we take a code length $N = 64$, information length $K = 57$ polar code as an example.

TABLE I
SIMULATION ASSUMPTIONS

| | |
|---|---|
| Channel | AWGN Channel |
| Modulation | QPSK |
| Info. Block length (=$K$ bits with out CRC) | $K = 8 : 1 : \min?(200, K_{\max,N})$, where $K_{\max,N} = \lfloor 5N/6 - crc \rfloor$; $K = \min?(200, K_{\max,N}) : 24 : K_{\max,N})$, where $K_{\max,N} = \lfloor 5N/6 - crc \rfloor$ excluding any code rates below $1/8$ |
| Codeword length (=$N$) | $\{64, 128, 256, 512, 1024\}$ |
| Decoding algorithm | List-X with LLR-based min-sum |
| List sizes | 1 and 16 (pruned to 8 best paths for CRC check) |
| Number of CRC bits (= $crc$) | 19 bits (0b1010001010110111 1001 where the last bit is d19) |

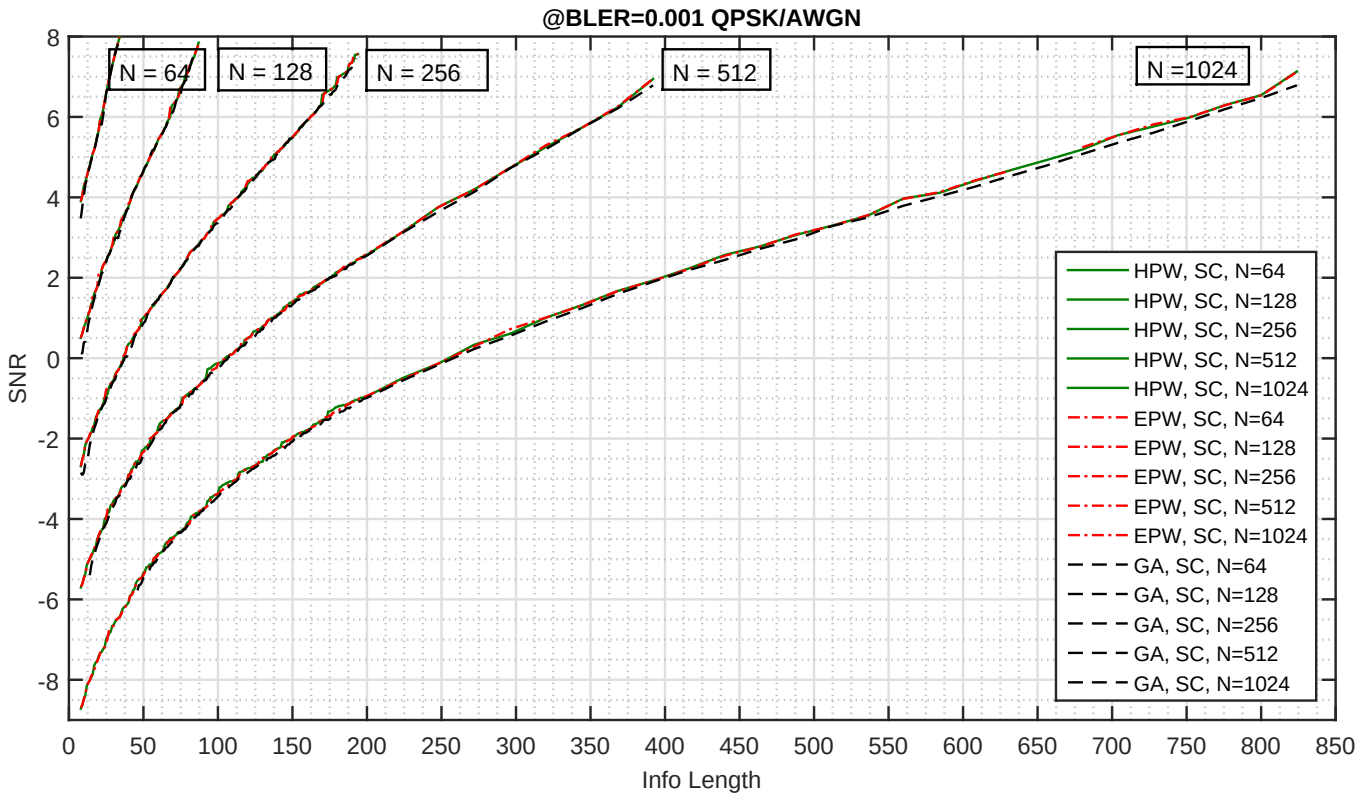

Fig. 3. The performance comparison between GA, HPW and EPW under SC decoder to achieve $10^{-3}$ BLER.

Through greedy searching of the construction SNR of GA method, the polar code construction with frozen channel set $\mathcal{F}_{\text{GA}} = \{0\ 1\ 2\ 4\ 8\ 16\ \mathbf{32}\}$ has better performance over all. However, the PW method with base $\beta = 2^{1/4}$ gives the frozen channel set as $\mathcal{F}_{\text{PW}} = \{0\ 1\ 2\ \mathbf{3}\ 4\ 8\ 16\}$. GA and PW treat sub-channel 3 and 32 in opposite reliability order.

Fig. 2 illustrated the calculated PW of each sub-channel for a 64 codeword length polar code. The PW pattern owns a quasiperiodic characteristic with a period of 4. Sub-channel 3 is on the peak of the 1st period and sub-channel 32 is on the valley of the 9th period. They are in different period, but have the PW value close to each other. The reliability order of sub-channel 3 and 32 is mistakenly flipped by PW method as $W_{32} = 2.378 > W_3 = 2.189$. Using a single base to present the PW cannot provide accurate enough information to sort the reliability order for all the sub-channels. Additional information should be added into the PW value to improve the quality of the reliability ordering and to cover the needs of various list size decoders.

Without loss of generality, we developed higher order PW (HPW) and extended PW (EPW) methods, as two examples, by generalizing PW methods to provide a more accurate reliability ordering.

### A. Higher Order Polarization Weight Method

Look back to (4), there is only one base $\beta$ in the summation. HPW method bring in new representation bases in the form of high-order $\beta$. For example, the 0th-order of the base is $\beta^{\frac{1}{4^0}}$ which equals to the original base $\beta$, the 1st-order of the base is $\beta^{\frac{1}{4^1}}$, and the $\xi$th-order of the base is $\beta^{\frac{1}{4^\xi}}$. Then, the representation is based on a summation of different bases. Assuming $\xi \in \Xi$ and $\Xi \subset \mathbb{Z}$, the HPW method can be presented as

$$W_i = \sum_{j=0}^{n-1} \sum_{\xi \in \Xi} B_j \times \beta^{\frac{1}{4^\xi}}. \tag{5}$$

For a higher-order base, the difference of different binary bits contribution to the corresponding base integration is compressed. Thus, HPW could describe the slight PW variations among each channels with the help of additional higher-order bases and provide a locally refined sub-channel sequence order. For implementation and description simplicity, we take HPW with 0th-order and 1st-order bases, $\Xi = \{0, 1\}$, as an example:

$$W_i = \sum_{j=0}^{n-1} B_j \times \left(\beta^j + \frac{1}{4}\beta^{\frac{1}{4}j}\right). \tag{6}$$

For HPW in (6), $W_{32} = 5.120 < W_3 = 5.733$, which has the same reliability ordering with GA. At the same time, Eq. (6) still hold the calculation simplicity of PW and the performance will be analyzed in the simulation section.

### B. Extended Polarization Weight Method

For the HPW, the new bases are only in the higher-order forms of the base $\beta$ in (4). The flexibility of the weight representation are still under the constrained of 0th-order base $\beta$. Extending the single base $\beta$ with a more general base $b$ is another way to achieve a refined representation of the sequence order with more flexibility. After the extension, the EPW can be written as

$$W_i = \sum_{j=0}^{n-1} B_j \times \left(\beta^j + a \times b^j\right), \tag{7}$$

where $a$ is the weight factor of base $b$. The new base $b$ in (7) provides a new degree of freedom to describe the PW relationship between each subchannels. The PW family formulas (4) - (7), mentioned above, all own the symmetry property. The ordering sequence of the first half are same to the second half, i.e. $\mathbf{q}_0^{N/2-1} = \mathbf{q}_{N/2}^{N-1} - N/2$. To further expand the capability of PW, a new factor called symmetry breaking point $B_c$ is brought into the formula to break the symmetry within a period of $2^{c+1}$. Then the ordering sequence of the first half and the second of $\mathbf{q}_{2^{c+1}k}^{2^{c+1}(k+1)-1}$ will not be the same, where $k = \left[0, 1, 2, \ldots, N/(2^{c+1}) - 1\right]$. If with three new general

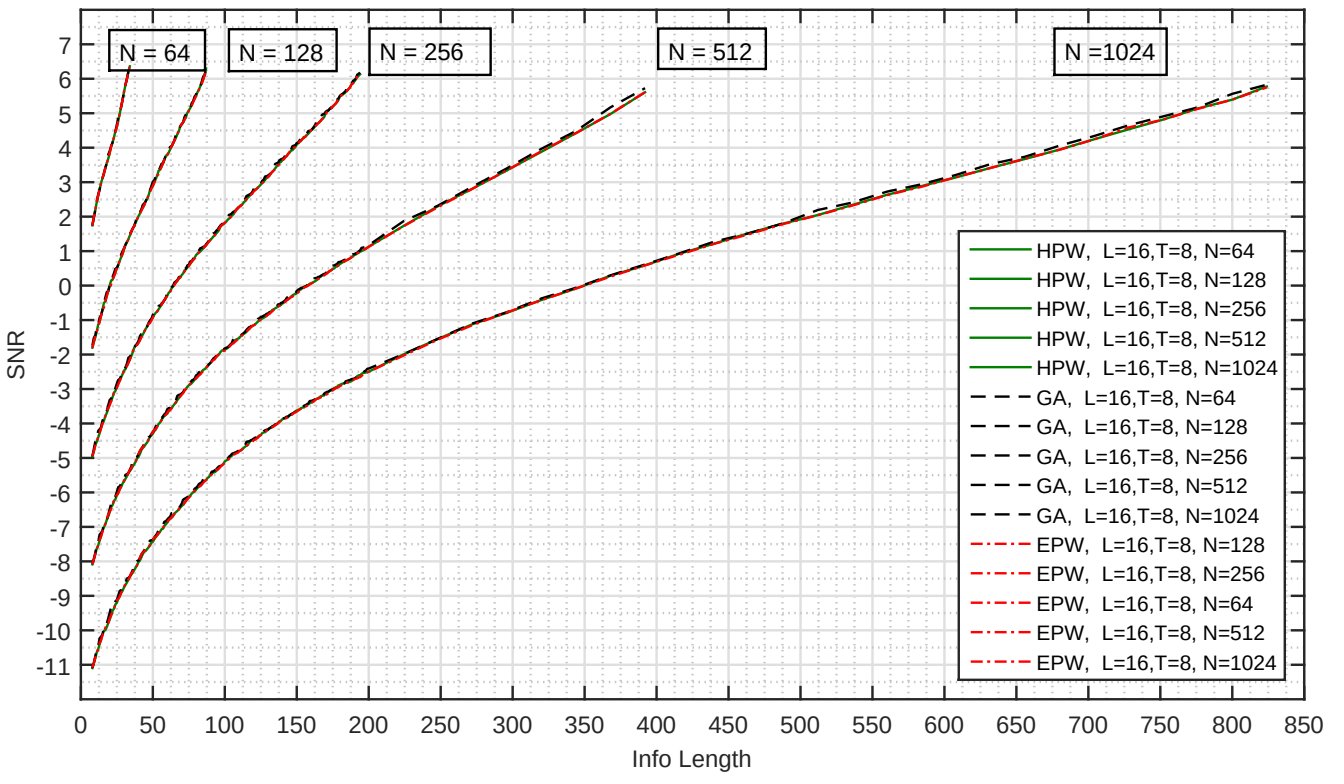

Fig. 4. The performance comparison between GA, HPW and EPW under $L = 16, T = 8$ decoder to achieve $10^{-3}$ BLER.

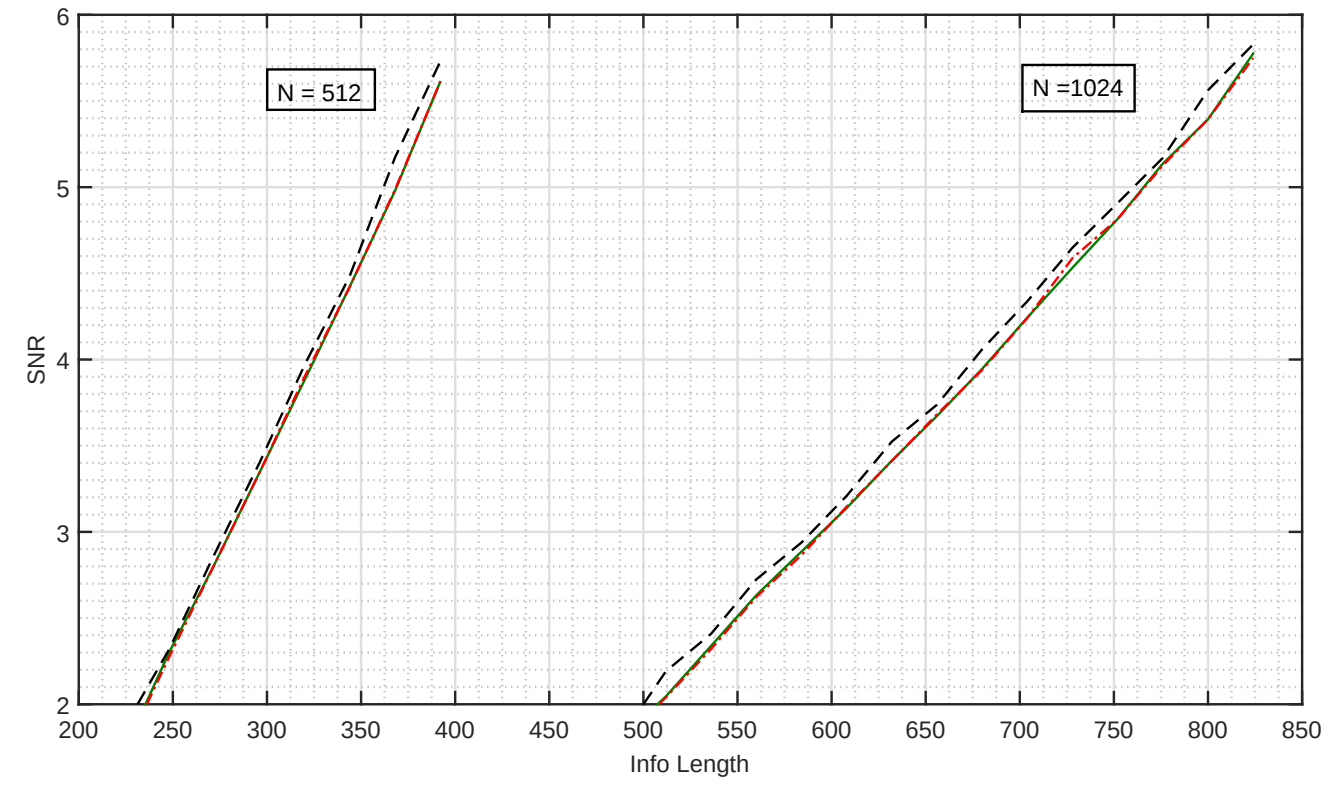

Fig. 5. The performance comparison between GA, HPW and EPW under $L = 16, T = 8$ decoder to achieve $10^{-3}$ BLER after zoom in.

bases and two symmetry breaking points, the EPW can be written as

$$W_i = \sum_{j=0}^{n-1} B_j \times \left(\beta^j + a \times b^j + B_c \times d \times e^j + B_f \times g \times h^j\right), \tag{8}$$

where $b$, $e$ and $h$ are extend bases, $B_c$ and $B_f$ are symmetry breaking points, and $a$, $d$ and $g$ are weight factors of each base.

The PW family sequences all own the nested characteristic, which means a sequence $\mathbf{q}_0^{N-1}$ of any codeword length $N < N_{\max}$ can be extracted from $\mathbf{q}_0^{N_{\max}-1}$ by selecting the indices where $q_i < N$. Obviously, the ordered index sequence $\mathbf{q}$, calculated by PW family methods, owns the nested property. The different codeword length only changes the integral upper limit $n-1$ in (4) - (8). Thus, a PW $W_i$ with $i \leq N-1$ is the same for codeword length $N$ or $N_{\max}$, and so does the ordered index sequence $\mathbf{q}_0^{N-1}$.

The mathematical simplicity of PW family methods make them easy to implement in chip. They can either be prestored in chip as a single sequence which covers $N_{\max}$ or calculated on-line with simple operations. They are fast polar code construction methods for various list size decoders.

## IV. Simulation

In this section, the performance of the new PW family member EPW and HPW are evaluated through numerical simulations. For HPW, the parameters in (6) are chosen. To evaluate the performance of EPW, we assume a set of parameters in (8) as an instance, then the EPW can be rewritten as

$$\begin{aligned} W_i = \sum_{j=0}^{n-1} B_j \times (&1.1892^j + 0.2210 \times 0.9889^j - B_8 \times 0.0371 \\ &\times 0.5759^j - B_7 \times 0.0470 \times 0.4433^j). \end{aligned} \tag{9}$$

The simulation assumptions are listed in Tab. I. The simulation cases cover different information block lengths in fine granularity for five codeword lengths (equal to $2^n$, where $n = 6, 7, 8, 9$ and $10$) and two typical list sizes of a LLR-based min-sum SC decoder. To present the simulation results clearly, we focus on evaluating the performance of the SNR to achieve $10^{-3}$ BLER by each construction method. The polar code construction method and the corresponding CRC polynomial are also listed in Tab. I. Here, the list-16 decoder are constrained to check the first 8 paths with the CRC detector in the survived 16 paths to secure a 16 bit CRC false alarm rate (FAR) ability [13].

For all the simulation cases, we simulated as many blocks as the accumulated error block number reached 2000 per SNR point to obtain a stable results. The SNR step is restricted to 0.1 dB to accurately estimate the SNR of achieving $10^{-3}$ BLER through linear interpolation. The performance of HPW and EPW are firstly evaluated under the SC decoder in company with GA, and the results are compared in Fig. 3. GA has almost the same precession as DE in binary input Gaussian channel under SC decoder, thus GA presents the lower bound of the required SNR for any sequence in this comparison. This figure illustrates that both HPW and EPW require similar SNRs with GA to achieve $10^{-3}$ BLER, especially when the codeword length is less than or equal to 512. For the 1024 codeword length cases, the required SNRs of EPW and HPW to achieve $10^{-3}$ BLER are not completely as low as GA when the information length $K$ is larger than 550. The performance of EPW and HPW on these cases should be continuously refined in the future work. Then, the performance are evaluated under a list-16 decoder and the comparison results are illustrated in Fig. 4. As the list size goes up, GA is no longer a competitive construction method. In Fig. 4, it is hard to tell the performance difference between HPW and EPW. It seems like both HPW and EPW require lower SNRs than GA to achieve $10^{-3}$ BLER. After focusing on the larger information length part of Fig. 4, the performance difference appears clearly in Fig. 5. The required SNRs of HPW and EPW still stick to each other and have almost the same value. Both HPW and EPW require lower SNRs than GA, especially as the information length $K$ goes longer. GA is designed to optimize the SC decoder performance, thus the designed SNRs for SC decoder do not work well under a list decoder.

The above simulation results indicate that both higher-order bases and extended bases bring in performance enhancement. They make the PW family method robust against different decoder list sizes, different information block lengths and different codeword lengths. The new bases only introduce small computational complexity in the summation, while still maintaining the mathematical simplicity of the PW family equation.

## V. Conclusion

PW family methods are the simple and neat formula defined methods to effortlessly construct polar code without considering channel conditions. The simulation results indicate that both the EPW and HPW methods have robust performance under various information lengths, codeword lengths, and decoder list sizes. The above mentioned HPW and EPW methods open a window to fast construct polar code for various list size decoders with a single sequence.